\documentclass{article}
\usepackage{authblk}
\usepackage[utf8]{inputenc}
\usepackage{graphicx}
\usepackage{hyperref}
\usepackage[sorting=none]{biblatex}
\usepackage{makecell}

\title{Is good old GRAPPA dead?}
\date{November 2020}

\author[1]{Zaccharie Ramzi}
\author[2]{Alexandre Vignaud}
\author[3]{Jean-Luc Starck}
\author[4]{Philippe Ciuciu}
\affil[1]{CEA (Neurospin \& Cosmostat), Inria (Parietal)}
\affil[2]{UNIRS , NeuroSpin, I2BM, DSV, CEA-Saclay, France}
\affil[3]{AIM, CEA, CNRS, Université Paris-Saclay, Université Paris Diderot, Sorbonne Paris Cité}
\affil[4]{Neurospin, Inria (Parietal)}

\def\qualifigsep{.1mm}
\begin{document}

\maketitle

\section{Synopsis}
We perform a qualitative analysis of performance of XPDNet, a state-of-the-art deep learning approach for MRI reconstruction, compared to GRAPPA, a classical approach.
We do this in multiple settings, in particular testing the robustness of the XPDNet to unseen settings, and show that the XPDNet can to some degree generalize well.

\section{Main findings}
XPDNet, a state-of-the-art deep learning approach for MRI reconstruction, can generalize well when compared to GRAPPA on unseen settings.

\section{Introduction}
Deep Learning has recently overwhelmed the field of MR image reconstruction from under-sampled data.
From AUTOMAP~\cite{Zhu2018} to the more recent unrolled neural networks~\cite{Schlemper2018, Hammernik2018LearningData, Sriram2020End-to-EndReconstruction, Ramzi2020XPDNetChallenge}, all of them showed the great potential of deep learning for the field, achieving ground-breaking results in quantitative metrics such as PSNR and SSIM for high acceleration factors.

However, there is a lack of comparison of these approaches to the current gold standards for periodic under-sampling, typically implemented in the MRI scanners.
In this paper, we compare a state-of-the-art approach, the XPDNet~\cite{Ramzi2020XPDNetChallenge}, to GRAPPA~(Generalized Autocalibrating Partially Parallel Acquisitions)~\cite{Griswold2002GeneralizedGRAPPA} on the task of reconstructing periodically under-sampled MR images in different qualitative settings.
GRAPPA is used in all the Siemens scanners, the most distributed in the world, as the default method for image reconstruction in the case of periodic under-sampling (and similar approaches for other manufacturers).
However, very few deep learning papers for MRI reconstruction include a comparison to GRAPPA.


\section{Methods}
\paragraph{Network.}
The XPDNet is a type of unrolled network that secured the second place in the 2020 fastMRI brain reconstruction challenge\cite{Muckley2020State-of-the-artChallenge}.
Very basically, it unrolls the PDHG~\cite{Chambolle2011AImaging} algorithm using an Multi-level Wavelet CNN (MWCNN)~\cite{Liu2018Multi-levelRestoration} as the learned proximity operator.
It has 25 unrolled iterations, and also features a sensitivity maps refinement module.
Two networks were trained for acceleration factors 4 and 8, using retrospectively under-sampled data from the fastMRI dataset~\cite{Zbontar} with equidistant Cartesian masks\footnote{\href{https://github.com/facebookresearch/fastMRI/issues/54}{facebookresearch/fastMRI/issues/54}}.
We chose to use non fine-tuned versions of the networks (i.e. trained on the four available imaging contrasts).

\paragraph{GRAPPA.}
We used the vanilla version of GRAPPA without noise handling.
We use kernels that span 5 points in the readout direction and 2 in the phase direction.
We manually set the regularisation parameter $\lambda>0$ to obtain the best compromise between quantitative and qualitative evaluation, therefore biasing the analysis towards GRAPPA\footnote{\href{https://www.youtube.com/watch?v=PngT6chFy6c&t=623s}{youtube.com/watch?v=PngT6chFy6c}}.
We leave the analysis of a smart setting of $\lambda$ for future works.

\paragraph{Data.}
We used 3 data sets to perform our comparison on:
\begin{itemize}
    \item a brain slice from the fastMRI validation data set~\cite{Zbontar} (the state-of-the-art network was trained on the training data set), with T2 contrast, {\bf retrospectively} under-sampled at acceleration factors 4 and 8;
    \item a brain slice acquired at a different resolution ($0.25mm \times 0.25mm$) using a different magnetic field strength~(7T), orientation and acceleration factor than the fastMRI brain data set and featuring the cerebellum~(not present in the fastMRI brain data set), with T2 contrast, {\bf prospectively} under-sampled at acceleration factor 2~--~this allows us to test the robustness of the network to somewhat unseen settings~\cite{Marrakchi-Kacem2016RobustChoices};
    \item a NIST phantom, {\bf prospectively} under-sampled at acceleration factor 8, acquired at 3T with 64 coils and a matrix size of $256 \times 256$.
\end{itemize}
All the data is periodically under-sampled with an Autocalibration Signal (ACS).

\paragraph{Code.}
We use the code linked in \cite{Ramzi2020XPDNetChallenge} for the network\footnote{\href{https://github.com/zaccharieramzi/fastmri-reproducible-benchmark}{github.com/zaccharieramzi/fastmri-reproducible-benchmark}}.
We used our own implementation of GRAPPA with a TensorFlow backend\footnote{\href{https://github.com/zaccharieramzi/grappa}{github.com/zaccharieramzi/grappa}}.

\section{Results}

\begin{figure*}[h]
\begin{center}
\hspace*{-3.5cm}\begin{tabular}{c@{\hspace*{\qualifigsep}}c@{\hspace*{\qualifigsep}}c}
{\bf Reference} & \makecell{{\bf GRAPPA} \\ PSNR: 35.83 \\ SSIM: 0.8784} & \makecell{{\bf XPDNet} \\ PSNR: 41.96 \\ SSIM: 0.9791} \\
\includegraphics[scale=0.35]{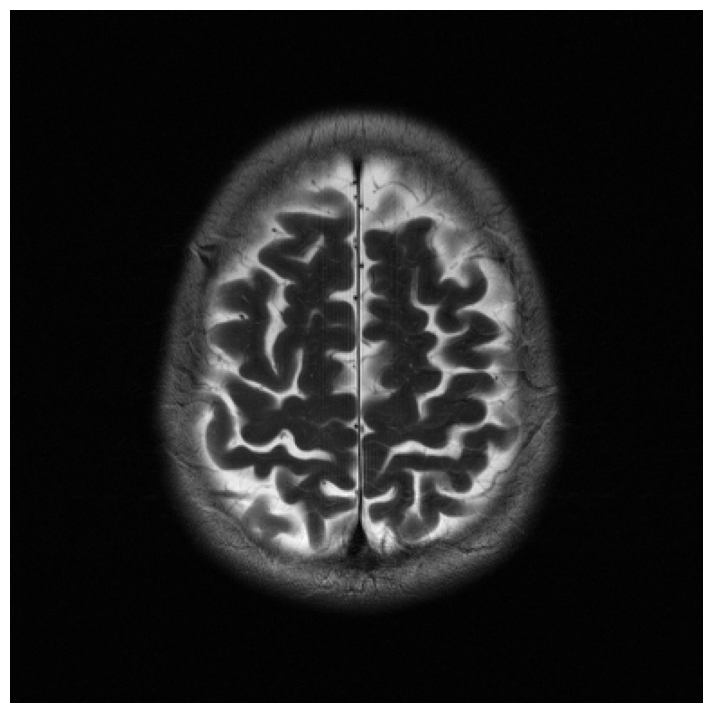}&
\includegraphics[scale=0.35]{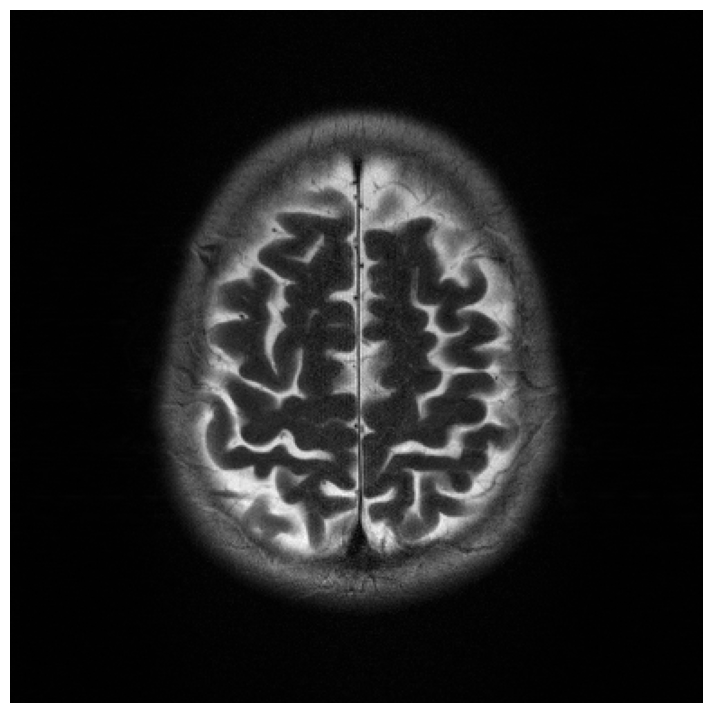}&
\includegraphics[scale=0.35]{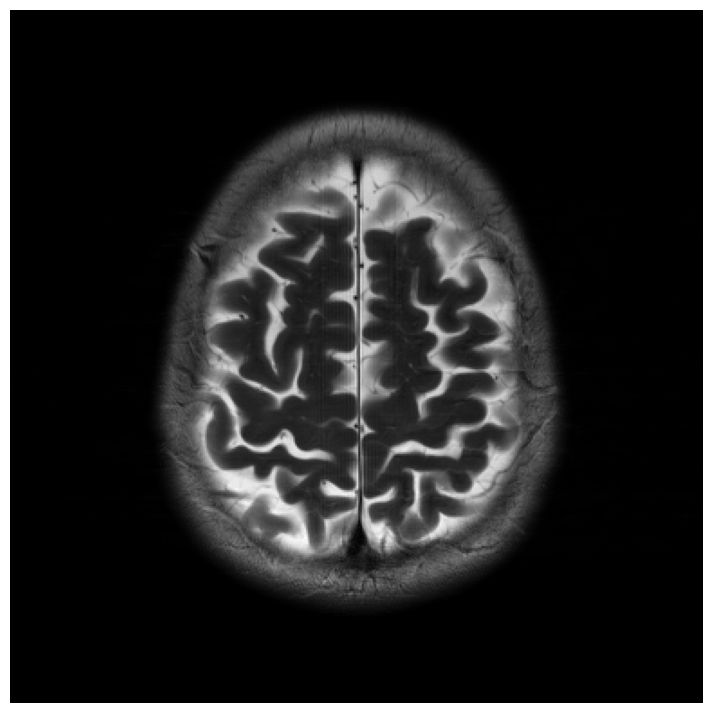}\\
&
\includegraphics[scale=0.35]{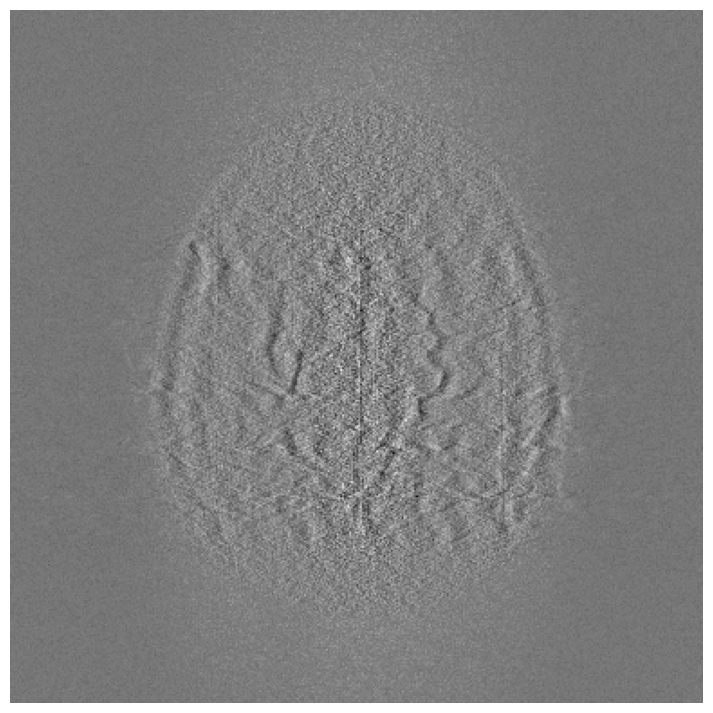}&
\includegraphics[scale=0.35]{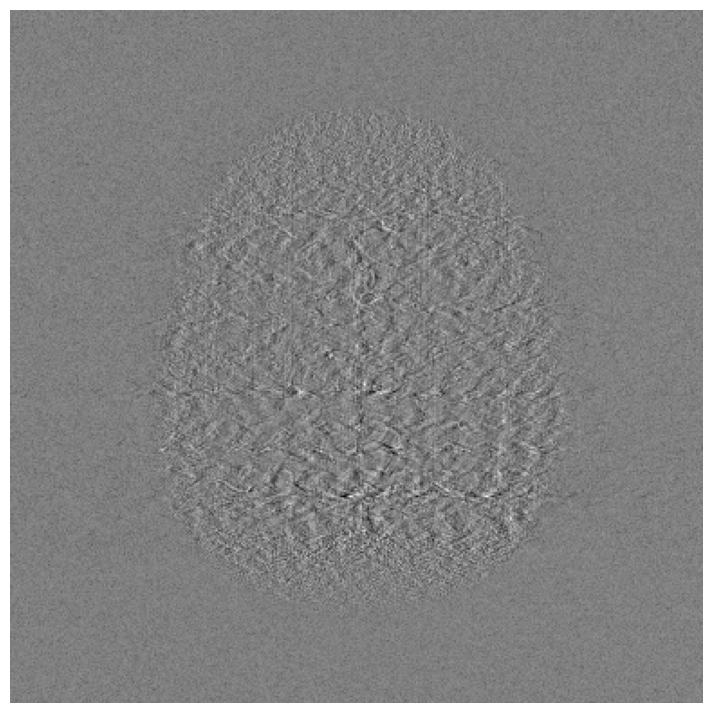}
\end{tabular}
\caption{Magnitude reconstruction results for a specific fastMRI slice at acceleration factor 4. The top row represents the reconstruction using the different methods, while the bottom one represents the error when compared to the reference. \label{fig:fastmri-r4}}
\end{center}
\end{figure*}

\begin{figure*}[h]
\begin{center}
\hspace*{-3.5cm}\begin{tabular}{c@{\hspace*{\qualifigsep}}c@{\hspace*{\qualifigsep}}c}
{\bf Reference} & \makecell{{\bf GRAPPA} \\ PSNR: 26.18 \\ SSIM: 0.7704} & \makecell{{\bf XPDNet} \\ PSNR: 36.82 \\ SSIM: 0.9626} \\
\includegraphics[scale=0.35]{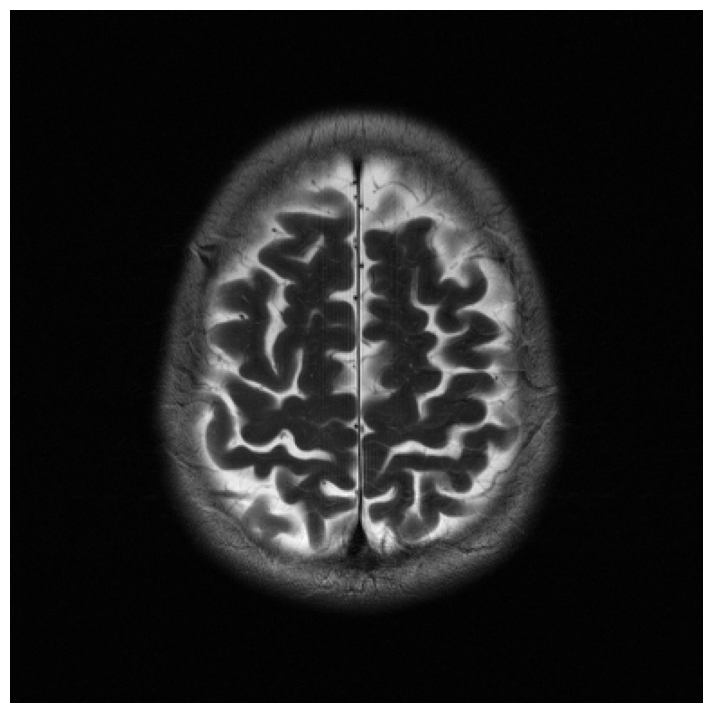}&
\includegraphics[scale=0.35]{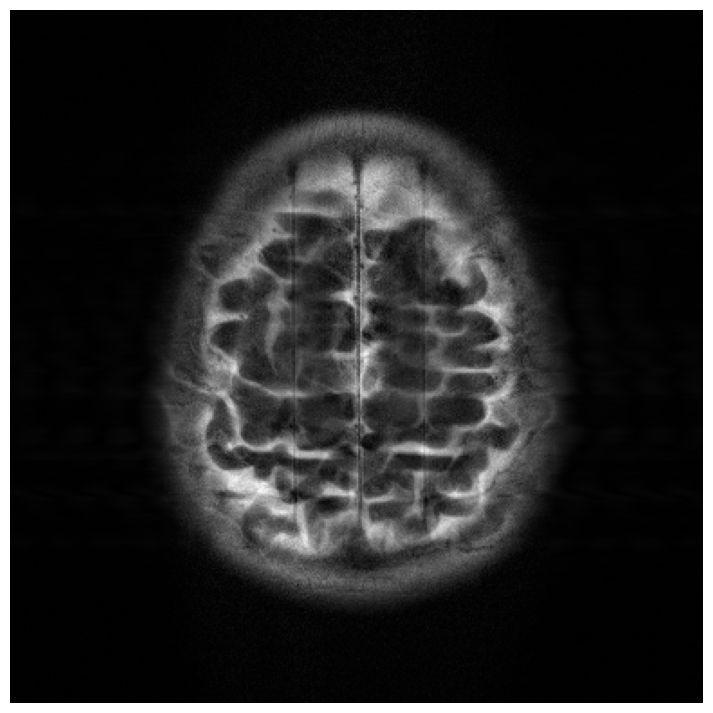}&
\includegraphics[scale=0.35]{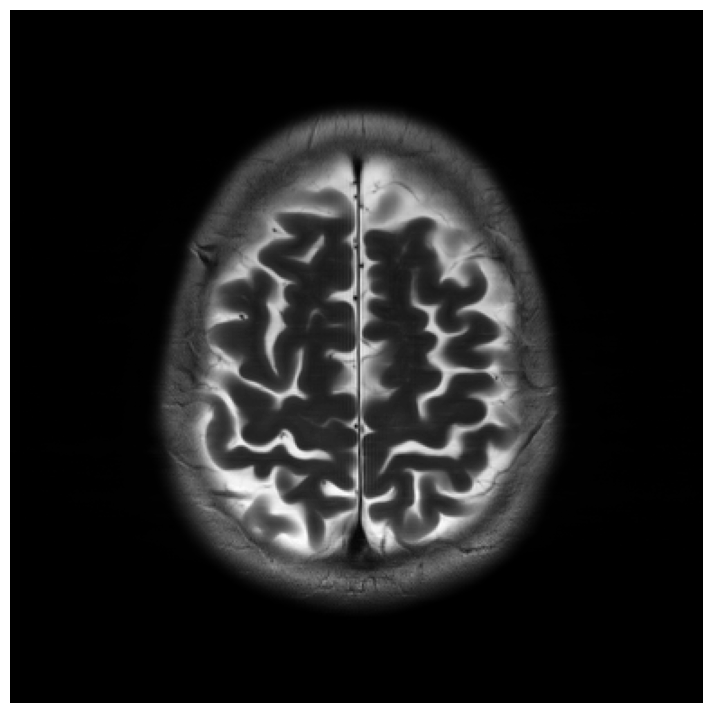}\\
&
\includegraphics[scale=0.35]{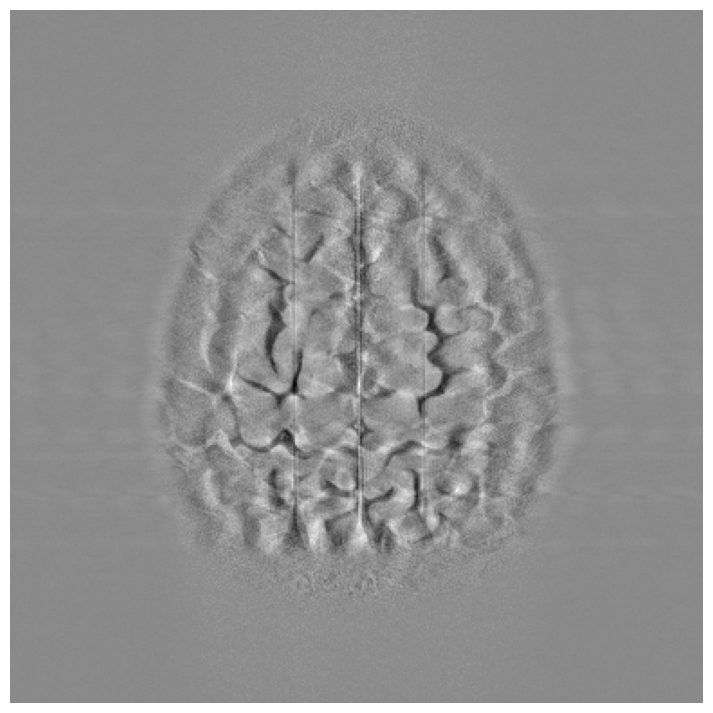}&
\includegraphics[scale=0.35]{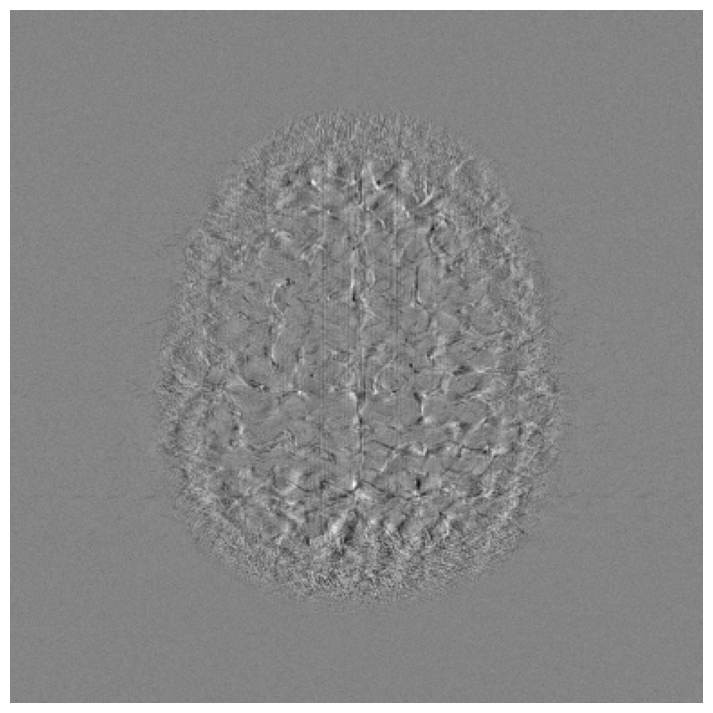}
\end{tabular}
\caption{Magnitude reconstruction results for a specific fastMRI slice at acceleration factor 8. The top row represents the reconstruction using the different methods, while the bottom row represents the error when compared to the reference. \label{fig:fastmri-r8}}
\end{center}
\end{figure*}

\paragraph{On the fastMRI brain slice.}
At acceleration factor 4, the quantitative results seem to show that the XPDNet has an overwhelmingly better image quality than GRAPPA. 
However, upon visual inspection of the images available in Fig.~\ref{fig:fastmri-r4}, we see that the image reconstructed by GRAPPA is only degraded by some noise not deteriorating its interpretability.
At acceleration factor 8, the quantitative metrics once again show a clear advantage of the XPDNet over GRAPPA.
This time, it is clearly confirmed by the visual inspection of the images presented in Fig.~\ref{fig:fastmri-r8}.

\begin{figure*}[h]
\begin{center}
\hspace*{-4.cm}\begin{tabular}{c@{\hspace*{\qualifigsep}}c}
\makecell{{\bf GRAPPA}} & \makecell{{\bf XPDNet}} \\
\includegraphics[scale=0.55]{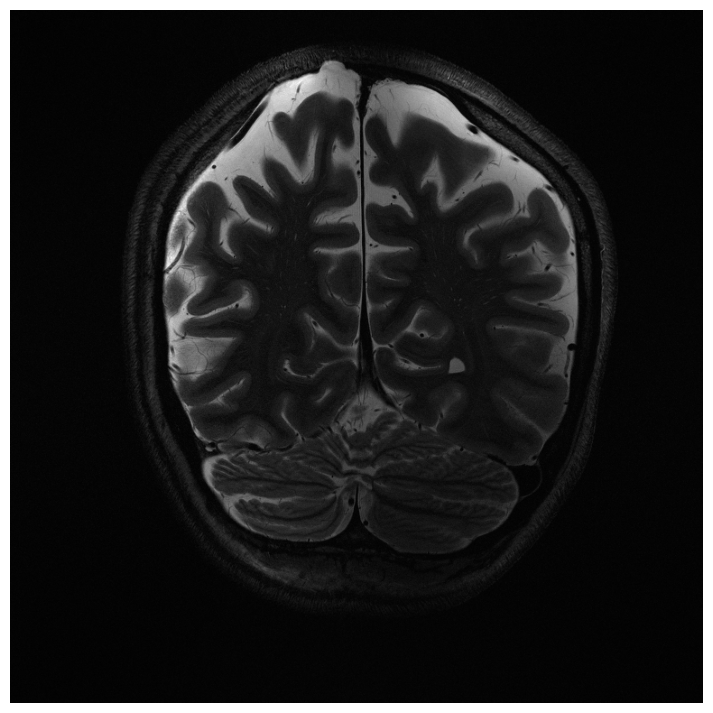}&
\includegraphics[scale=0.55]{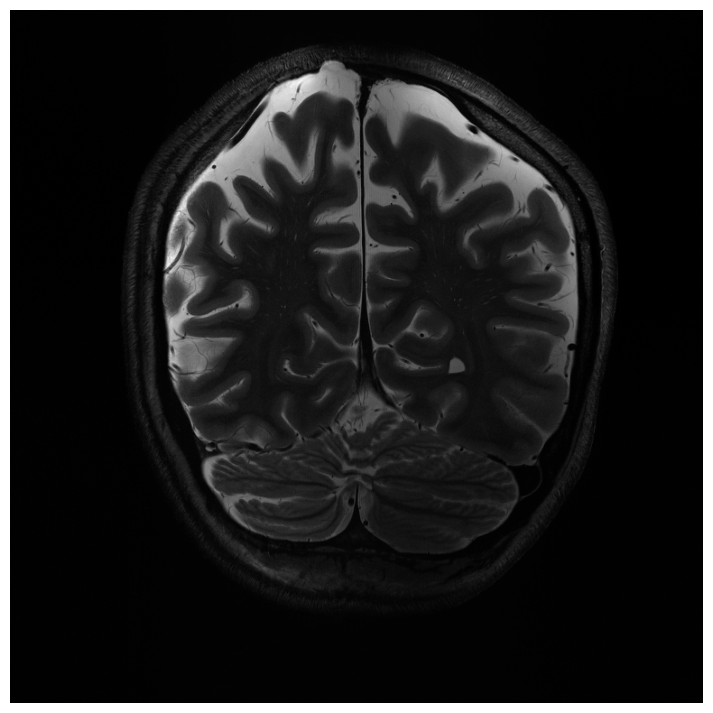}\\
\includegraphics[scale=0.55]{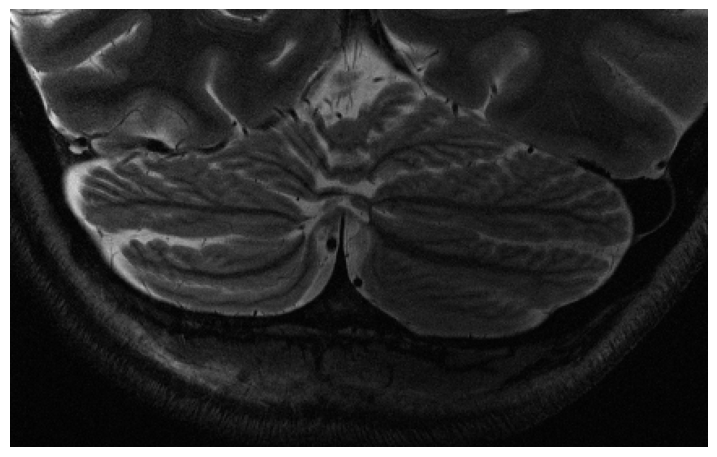}&
\includegraphics[scale=0.55]{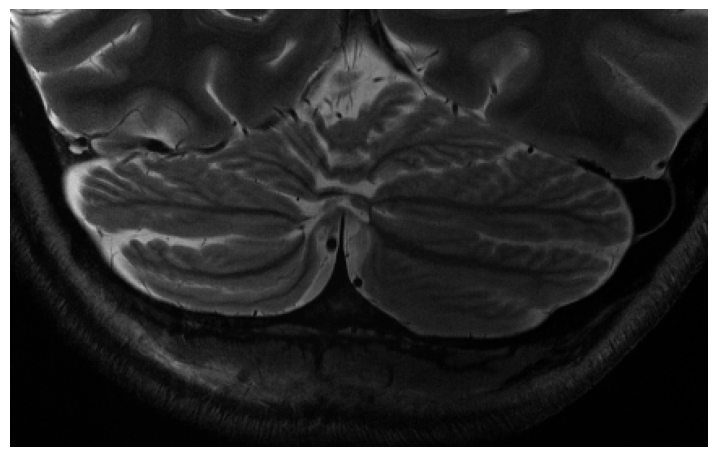}
\end{tabular}
\caption{Magnitude reconstruction results for a brain acquired at acceleration factor 2, contrast T2, and field strength of 7T. The top row represents the reconstruction using the different methods, while the bottom one represents a zoom in the cerebellum region, an anatomical feature that was not present in the XPDNet training set. \label{fig:brain-7t}}
\end{center}
\end{figure*}

\paragraph{On the out-of-distribution brain slice.}
The image reconstructed with the XPDNet shows some faint smoothing in the cerebellum as shown in the bottom row of Fig.~\ref{fig:brain-7t}.
However, the overall image is artifacts-free and very difficult to distinguish from the GRAPPA-reconstructed one.

\begin{figure*}[h]
\begin{center}
\hspace*{-3.5cm}\begin{tabular}{c@{\hspace*{\qualifigsep}}c@{\hspace*{\qualifigsep}}c}
{\bf Reference} & \makecell{{\bf GRAPPA} \\ PSNR: 24.59 \\ SSIM: 0.7673} & \makecell{{\bf XPDNet} \\ PSNR: 18.40 \\ SSIM: 0.6328} \\
\includegraphics[scale=0.35]{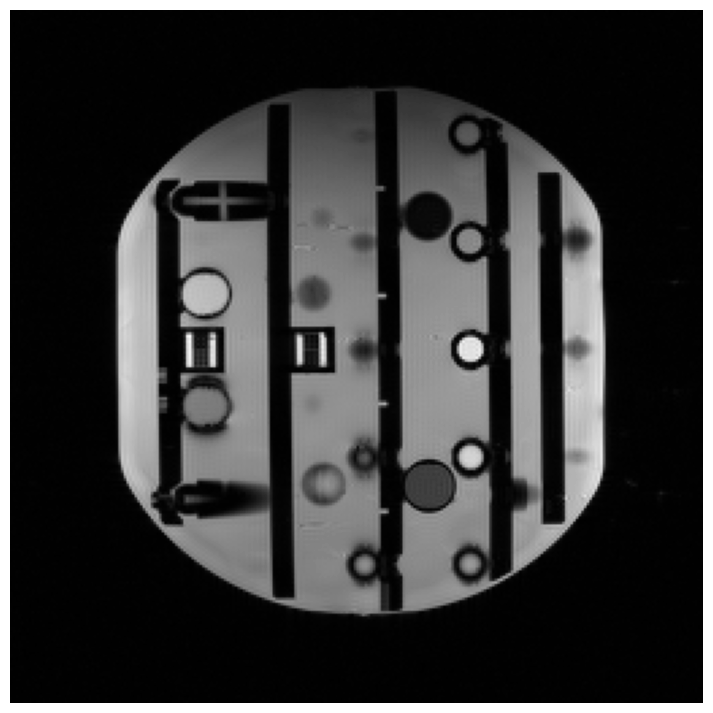}&
\includegraphics[scale=0.35]{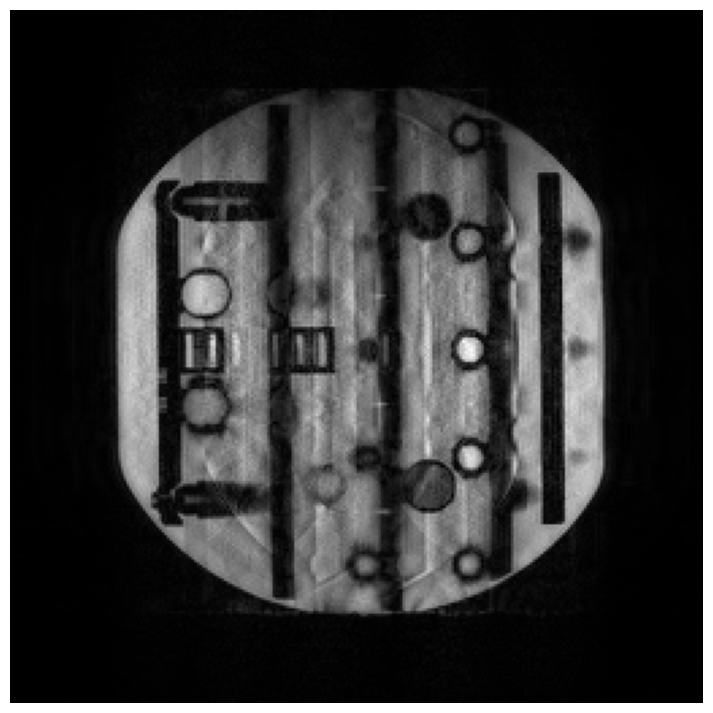}&
\includegraphics[scale=0.35]{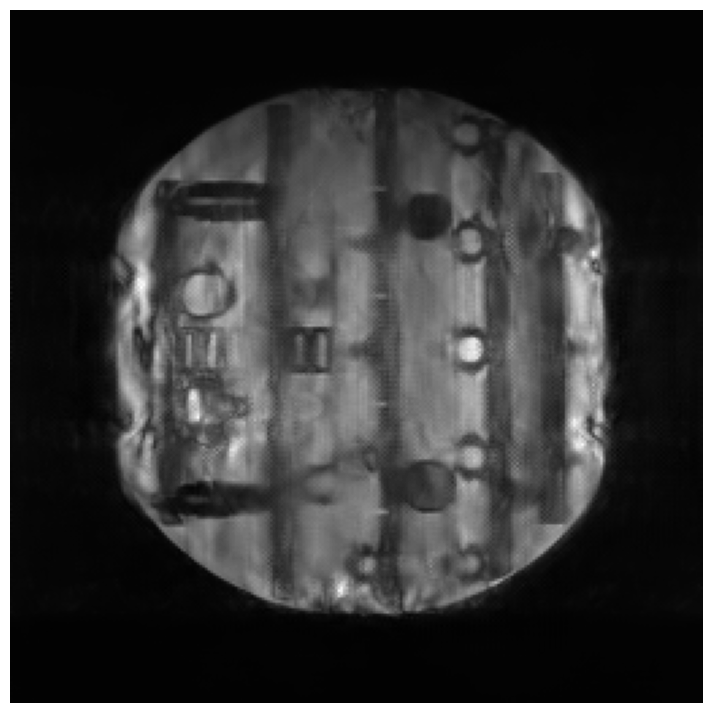}\\
&
\includegraphics[scale=0.35]{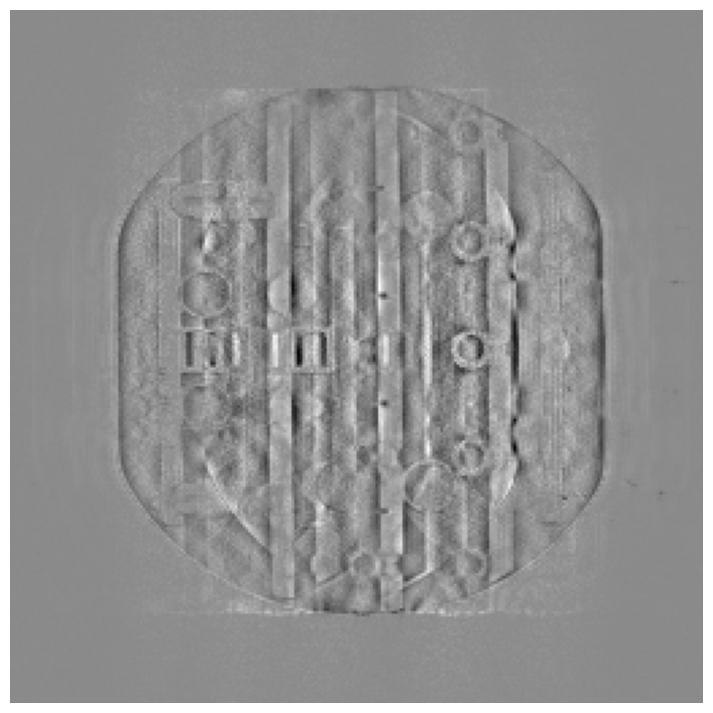}&
\includegraphics[scale=0.35]{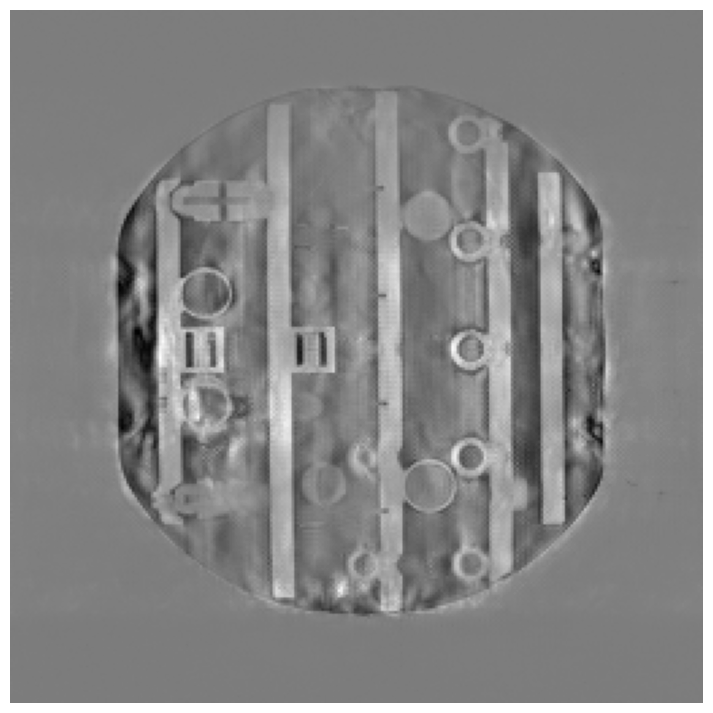}
\end{tabular}
\caption{Magnitude reconstruction results for a phantom acquired at acceleration factor 8. The top row represents the reconstruction using the different methods, while the bottom one represents the error when compared to the reference. \label{fig:phantom}}
\end{center}
\end{figure*}

\paragraph{On the NIST phantom.}
The phantom reconstructed using GRAPPA and XPDNet at acceleration factor 8 are very poor.
For GRAPPA, the noise completely obfuscates the signal while for XPDNet the artifacts are present everywhere as can be seen in Fig.~\ref{fig:phantom}.

\section{Conclusion and Discussion}

The Deep Learning techniques seem ready for a substitution test at acceleration factor 4, however, they do not seem to provide an overwhelming advantage over GRAPPA visually.
The acceleration factor of 8 looks like an attainable target, and it would drastically improve the image quality when compared to GRAPPA, even when using the latest noise handling techniques.
We also showed that the XPDNet is robust enough to be adapted to settings relatively different from the training distribution. 
However, if trained on brains a network can not reconstruct objects that are too dissimilar, like phantoms.
We conclude that it is therefore important to test visually the results of a reconstruction network at low acceleration factors to measure the difference compared to GRAPPA, and that the high acceleration factors are currently the real target for Deep Learning.

This work demands other types of robustness and sanity tests such as (but not limited to) receiver array coil design, SNR level, contrasts, organs, and orientation.

\section{Figures}

\printbibliography

\end{document}